\documentclass{vldb}
\usepackage{appendix}
\usepackage{booktabs} % For formal tables
\usepackage{balance}
\usepackage{listings}
\usepackage{xcolor}
\usepackage{pbox}
\usepackage{verbatim}
\usepackage{graphicx}
\usepackage{url}
\usepackage{color}
\usepackage{algorithmic}
\usepackage{verbatim} 

\newcommand{\idebench}{IDEBench}

\vldbTitle{A Sample Proceedings of the VLDB Endowment Paper in LaTeX Format}
\vldbAuthors{P. Eichmann et al.}
\vldbDOI{https://doi.org/TBD}

\begin{document}

\title{IDEBench: A Benchmark for Interactive Data Exploration}

\author{
\begin{tabular}{cccc}\\
Philipp Eichmann \textsuperscript{1} & Carsten Binnig \textsuperscript{1,2} & Tim Kraska\textsuperscript{1,3} & Emanuel Zgraggen\textsuperscript{1}
\end{tabular}\\
\begin{tabular}{ccc}\\
\textsuperscript{1}\eaddfnt{Brown University, USA} &
\textsuperscript{2}\eaddfnt{TU Darmstadt, Germany} &
\textsuperscript{3}\eaddfnt{MIT CSAIL, USA} 
\end{tabular}
}\maketitle

\begin{abstract}
Existing benchmarks for analytical database systems such as TPC-DS and TPC-H are designed for static reporting scenarios. The main metric of these benchmarks is the performance of running individual SQL queries over a synthetic database.  
In this paper, we argue that such benchmarks are not suitable for evaluating database workloads originating from interactive data exploration (IDE) systems where most queries are ad-hoc, not based on predefined reports, and built incrementally. 

As a main contribution, we present a novel benchmark called \idebench{} that can be used to evaluate the performance of database systems for IDE workloads. 
As opposed to traditional benchmarks for analytical database systems, our goal is to provide more meaningful workloads and datasets that can be used to benchmark IDE query engines, with a particular focus on metrics that capture the trade-off between query performance and quality of the result.
As a second contribution, this paper evaluates and discusses the performance results of selected IDE query engines using our benchmark. 
The study includes two commercial systems, as well as two research prototypes (IDEA, approXimateDB/XDB), and one traditional analytical database system (MonetDB).
\end{abstract}

\section{Introduction}
\label{sec:introduction}

\textbf{Motivation:} 
There is an ever growing need for systems which allow data scientist of varying skills levels to interactively and often visually explore large structured data sets. 
Unfortunately, traditional analytical database systems, such as MonetDB\cite{monetdb} or SAP HANA \cite{saphana}, usually do not provide the main property these systems require:  sub-second response times for ad-hoc queries created by and for a visual interface \cite{dice2,dice,idea}. 
As a result many existing interactive data exploration (IDE) systems started to create their own specialized query execution engines. 
For example, Tableau has its own SQL engine \cite{tableau}, Vizdom \cite{vizdom} has the Interactive Data Exploration Accelerator IDEA \cite{idea} as a seperate execution engine, and imMens \cite{immens} uses a special cube-based representation and query engine.
In addition, there has also been  attempts to build approximate query engines independent of the interface, such as SnappyData \cite{snappy} or DICE \cite{dice2,dice}, with the goal to support interactive data exploration. 
All those systems have in common that they aim to provide better support for IDE workloads often by taking advantage of the fact  that (1) users use visual tools to incrementally create more and more complex queries  and that (2) good approximate answers returned in seconds are better than precise answers in minutes. 

However, those systems widely differ in the techniques they use and the trade-offs they make. 
For example, IDEA  uses specialized data structures and on-demand creation of stratified samples, whereas DICE does speculative pre-com\-putation based on previous interactions.  
Furthermore, different engines implement different execution models (blocking, approximate, progressive), and might or might not re-use previously computed results \cite{idea, approxreuse, vistrees, tupleware}.
Additionally, IDE systems need to make a trade-off between the amount of pre-processing the system requires (i.e., how long does a user have to wait before they can explore a new data set) versus the quality of the first approximate answers. 
Even worse, different systems might have different trade-offs in regard to time vs. quality; some systems might aim to provide a first answer in 500ms, the general interactivity threshold \cite{liu2014effects},  other systems will only return a result when the quality of the answer reaches some threshold, whereas others aim for progressive results in which the quality of an answer improves as longer the user is willing to wait. 

This variety of goals and trade-offs make it extremely hard to determine which system is better for a particular data exploration task. 
For example, how much advantage do pre-computed stratified samples provide.
How much overhead do approximate query processing techniques introduce? When would a traditional system like MonetDB not simply outperform an approximate engine without having even approximate answers.
Or even more mundane, which of two approximate query engines is better. 

Unfortunately, traditional analytical benchmarks can ba\-rely be used to answer any of these questions. 
For instance, both TPC-H \cite{tpch} and TPC-DS \cite{tpcds} use the time per query as the main metric, thereby not taking into consideration that results might be approximate. 
Di Tria et al. \cite{di2017benchmark} developed an extension for TPC-H that takes the accuracy of approximates result into account, but does not factor in confidence interval in their metrics.
More importantly, these benchmarks assume a fixed upfront known workload of complex queries, whereas in IDE queries are ad-hoc and built incrementally. 
Finally, most existing analytical benchmarks assume that only one query at the time runs, whereas visual interfaces often create several queries, which run simultaneous. \\

\textbf{Contributions:} The first contribution of this paper is  a \textbf{novel benchmark called \idebench{} that can be used to evaluate the performance of database systems for IDE workloads} under realistic conditions in a standardized, automated, and re-producible way.
Unlike the static workloads of traditional benchmarks, \idebench{} measures the performance of IDE systems over the course of workloads that are closer to real data exploration workflows.

One of the key challenges in defining an IDE benchmark is to decide what a typical workflow constitutes.
We therefore build upon previous user studies \cite{idea,liu2014effects} and derived three common IDE browsing patterns ranging from independent browsing, where users investigate the distribution of attributes of a dataset and specify arbitrary filters, to more targeted scenarios, where users want to answer a specific question.
Yet, {\em our goal is NOT to simulate users}, which is arguably impossible. As Hawking puts it, ``Intelligence is the ability to adapt to change'' and we are far away from being able to simulate users, which can react intelligently to varying quality of query results, build or loose trust in a system based on answers, or come up with new conclusions or exploration paths based on particular insights. 
Similarly, our goal is not to benchmark the effectiveness of user interfaces or other visual components, such as visual recommendations \cite{seedb}. 

Rather our goal is to provide meaningful abstractions, which provides a first step towards benchmarking IDE query engines with a particular focus on metrics that capture the trade-off between query performance and quality of the result.
Furthermore, we created a highly customizable benchmarking framework, which
allows research groups to change different benchmarking settings to their envisioned user scenario while still enabling a high degree of reproducibility and comparability. 

The second contribution of this paper is a comprehensive study of running the benchmark on different database engines. 
The study includes two IDE engines as well as two research prototypes (IDEA, approXimateDB/XDB) and one traditional analytical database system (MonetDB). 
The benchmark code is available for download at \url{http://idebench.github.io}.

\textbf{Outline:} The remainder of the paper is organized as follows:
In Section \ref{sec:landscape}, we first reiterate the current landscape of IDE systems and their typical workload characteristics.
We then discuss shortcomings of existing IDE evaluation approaches and derive a set of requirements for an IDE benchmark in Section \ref{sec:requirements}.
Section \ref{sec:benchmark} presents the design of \idebench{} and discusses the workload and data generator, as well as the metrics and reporting requirements.
In Section \ref{sec:results} we present the results of running the benchmark on five database systems; two commercial systems, two research prototypes (IDEA, approXimateDB/XDB), and one traditional analytical database system (MonetDB).
Finally, we discuss the most important findings of the benchmark results and discuss future work in Section \ref{sec:discussion}.

\vspace{4ex}
\section{Interactive Data Exploration}
\label{sec:landscape}

\begin{figure*}
\hspace{-1.5ex}
\includegraphics[width=1.08\textwidth]{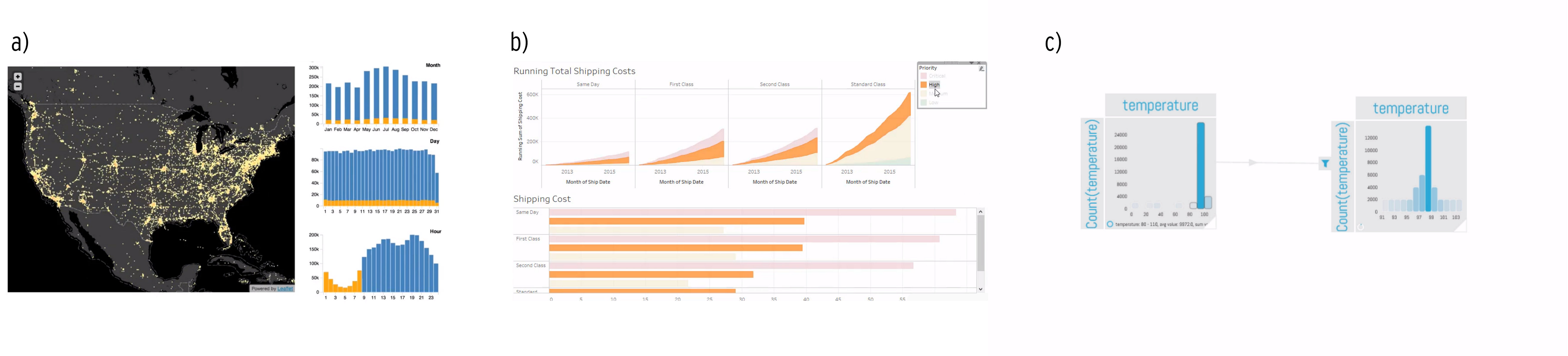}
\caption{Shows example visualizations in three different IDE frontends: a) imMens \protect\cite{immens} b) Tableau \protect\cite{tableauvideo} c) Vizdom \protect\cite{vizdomvideo}. All display binned plots, but in a and b links are implicit whereas in c links are explicitly drawn by the user (gray arrow). }
\label{fig:systems}
\end{figure*}

Based on findings in recent user studies \cite{idea,liu2014effects} we first describe a use case that anecdotally exemplifies common visual IDE frontends and then summarize typical workload characteristics that emerge for database systems that are triggered by these frontends.
We then provide an overview of IDE systems which are often used in practice, and discuss why existing evaluation approaches are not suitable to compare these systems.

\vspace{1pt}

\subsection{Use Case}
Imagine Jean: a research staff member at a major hospital. 
She wants to get an overview of the hospital's patient population and their health problems.
To do so, she looks at electronic health records from the past 20 years. 
Jean starts out by examining demographic information of patients and, for example, finds that patients ages are normally distributed. 
She then continues to look for interesting patterns in admission times and dates. 
Jean creates a query that shows the number of new admits per hour of the day. The result reveals that most admits are during business hours, but there is an interesting bump from 7 to 10pm. 
She filters down to admits coming from the emergency center and notices that most of the admits between 7 and 10pm are coming from there. 
Is this trend identical on all days of the week? 
She refines her query to only show the admits on weekends and sees that the previous evening bump now shifted towards 10 to 12pm. 
Who are these patients? Jean filters her previous age query by patients admitted on weekends between 10 and 12pm. She finds that patients ranging from 20 to 35 are over represented in this subset when compared to the overall age distribution. 
Now Jean wants to see which health problems are common among this sub-population.
She finds that head traumas are fairly frequent and decides to check with the administration if the hospital's duty rota accommodates for this by making sure a trauma specialist is on call during weekend nights. 

\subsection{IDE Workload Characteristics}
\label{workload_characteristics}

Based on this exemplary use case, we now discuss the most important workload characteristics that emerge for database systems.

\paragraph*{Aggregation Queries} 

IDE workloads are dominated by OLAP-style aggregation queries and typically follow the ``Visual Information Seeking Mantra'' \cite{shneiderman1996eyes} of ``Overview first, zoom and filter, then details-on-demand''.
Since the result of queries are typically visualized by an IDE frontend, most queries group the data by one or many attributes and apply aggregate functions to each group such as \texttt{AVG}, or \texttt{SUM}.

When dealing with large datasets, however, visualization systems commonly bin the data by some definition in order to compute aggregate results instead of just simply grouping on a set of attributes.
Binning can be found for a wide-range of visualizations such as histograms and their siblings, pie charts, choropleth maps, or bubble charts. 
Binned plots are omnipresent and featured in virtually any visualization software.
Figure \ref{fig:systems} illustrates this with screenshots of three different IDE frontends: imMens, Tableau and Vizdom.
Defining the binning behavior of an IDE frontend and thus for a benchmark, which includes the dimensionality of a bin (e.g., 1D for a histogram, 2D for a binned scatter plot), and the bin boundaries is not trivial.
When the distribution of data for a visualization is unknown, there are generally two methods how bin boundaries can be defined:
1) they can be specified by either choosing a pre-defined number of bins, which, for quantitative values requires a computation of the current minimum and maximum value of each bin.
2) they could be defined by choosing a interval based on a fixed bin width and a reference value.

\paragraph*{Incremental Query Building and Think Time}

As illustrated in the use case, IDE frontends are often used in an iterative process where queries are constructed ad-hoc and are refined incrementally.
For example, users typically start looking at all data and then narrow down the search to more interesting details. 

Furthermore, visualizations in IDE frontends are often linked together. 
Linking refers to setting the data source of a visualization (target) to the data source of another visualization (source).
When data of a source visualization is either filtered or selected, either the source and the target, or just the target visualization are forced to update.
This is illustrated in Figure \ref{fig:systems} where changing the selection in one visualization updates all other visualizations.
Linking on the query level often represents a join, since attributes of different tables might be connected that way.

Finally, user interactions are typically separated by a think-time, during which users analyze results and decide on what to do next.

\paragraph*{Multiple Concurrent Queries}

Many existing visual tools for IDE e.g. \cite{tableau,vizdom,immens} provide user interfaces to informally create and layout visualizations of different subsets of the data, apply (cross-) filtering, and to perform ``linking, brushing and zooming'' operations \cite{keim,visscientific}.
Typically, this allows users to look at different facets of a data set at the same time.

Consider Figure \ref{fig:systems}: each of the depicted applications \cite{tableau, immens, vizdom} displays linked visualizations that users can utilize to simultaneously brush or filter other visualizations.
Such links are either implicitly created by the application or database schema, or explicitly defined by the user.

In more abstract terms, dash-boards built by users using an IDE frontend can be seen as dependency graphs of visualization and filter objects.
Changing properties of either object may require all dependent visualization to update, which on the database-level leads to multiple concurrent queries per interaction.

\subsection{IDE Database Landscape}

Several commercial and academic database systems aim to support interactive data exploration workloads.
In the following we summarize this landscape through three categories and provide examples for each. 

\paragraph*{Analytical Database Systems} This category represents classic database systems that efficiently execute aggregate queries to completion and then return full result. 
This includes column-stores and main-memory systems such as MonetDB \cite{monetdb}, SAP HANA \cite{saphana}, Hyper \cite{hyper} as well as database management systems that are designed for online analytical processing (OLAP) type workloads ~\cite{chaudhuri1997overview}.
These systems cannot guarantee interactive response times on large data sets.
However, since they are often used as backends for many visual data analytics tools such as Tableau, we include them in our experiments as a baseline.

\paragraph*{Approximate Database Systems} 
This category of systems also targets aggregate queries.
Contrary to classic analytical database systems these tools either use offline or online sampling techniques to return an approximate answer without scanning the entire dataset. 
That way, these systems often can better guarantee interactive response times even on large data sets.
Examples in this category include AQUA \cite{acharya1999aqua}, Snappy Data, BlinkDB \cite{agarwal2013blinkdb}, and ApproXimate DB (XDB) \cite{li2016wander}. 
Individual systems support different ways of how users interact with them; some require users to set a desired result quality per query or a fixed time constraint. 

\paragraph*{Specialized Engines for IDE} 
Systems in this category represent specialized engines for IDE that often come with their own custom user interface. 
A prominent commercial example is the backend of Tableau~\cite{tableau} and its research predecessor Polaris~\cite{Polaris2002}, which uses a specialized engine for visual data analysis workloads.
imMens~\cite{immens} is another example that includes a specialized engine heavily leverages pre-computation over the possible query space to enable interactivity.
IDEA \cite{approxreuse} is a backend used for a pen-and-touch interface.
It uses online sampling techniques to progressively compute query results and push them to the user interface on request. 
DICE \cite{dice2} is a backend system optimized for exploratory cube analysis and leverages interaction delays (i.e., ``think-times'') and a user interaction model to predict future queries.
\section{Towards a new Benchmark}
\label{sec:requirements}
Recent position papers \cite{battle2009position,idebench} have advocated for an IDE benchmark.
In this section we discuss the scope of our work and the requirements we believe to be crucial for a benchmark for IDE backend systems.

\pagebreak

\subsection{Scope}
\label{scope}

Ideally workloads are close to the database workload generated by the behavior of real users in a variety of different data exploration scenarios.
While we acknowledge the breadth and richness of different IDE tasks and user interfaces (UI), this work does not attempt to model exact human behavior for its workloads.

Instead we follow a more pragmatic, UI-agnostic approach: we limit the scope of \idebench{} to common exploration patterns which can be translated to workloads on the database backend.
As previously discussed, we observed these patterns in a range of user studies (e.g., \cite{idea, zgraggen2014panoramicdata}), and empirically found them to be supported by many other modern IDE systems (see Figure \ref{fig:systems}).
To that end, \idebench{} focuses on aggregate queries on large datasets where queries are built and refined incrementally (separated by a think-time), result are (cross)-filtered between tables, and applying filters on linked visualizations can result in multiple concurrent queries.

Furthermore, in its current version \idebench{} targets data warehouse star schemas that are often used in analytical scenarios in both de-normalized and normalized form.
In future, we plan to extend the benchmark to more rich database schemata as well.

\subsection{Requirements}

In the following we discuss the workload, the metrics, and the data that a new benchmark such as \idebench{} should fulfill.

\paragraph*{Workload}

As discussed before, we believe that a new benchmark should provide a workload that could result from plausible user behaviours (see Section \ref{workload_characteristics}) where actions to build and modify a visualizations result in queries to an IDE backend.
As discussed before, we therefore require that the workload not only runs individual queries but resemble workload characteristics discussed in Section \ref{workload_characteristics} that can be triggered by user interacting with an visual IDE frontend.
This includes actions to create new visualizations which triggers new queries, actions that add filters to incrementally select a sub-populations, link visualizations where a single action can force multiple visualizations to update.
Furthermore, when queries in the benchmark are executed, there must be delays between queries triggered by consecutive user interactions, and the queries by the simulation must be aggregation queries with different parameterizations, i.e., different binning strategies and aggregate functions.
Lastly, it is crucial for workloads to be applicable to the wide range of database backends, as discussed in Section \ref{sec:landscape}.

\paragraph*{Metrics} 

Metrics used to evaluate the performance of an IDE system should capture their generative power for insights, and must be applicable to IDE systems which return exact or progressive/approximate results.
We believe that this can be reduced to two aspects of IDE backends.
(1) Speed: a recent study \cite{liu2014effects} shows that higher query latencies negatively affects users and their ability to derive insights from data.
(2) Quality of the results: while approximate and progressive systems are able to maintain low latencies, intermediate results may vary from the ground-truth, or could be returned with low confidence (large confidence intervals). 
Because there is a trade-off between speed and quality, we believe that the metrics for an IDE benchmark should reflect the quality of the results for a given interactivity/time requirement (quality after $x$ seconds).
Important here is that the metrics should also be applicable to classic analytical database systems, where the query results are always exact but the query execution may be slow since results are only returned upon query completion.

\paragraph*{Data} 

An IDE benchmark should support different datasets with different scale factors. 
As some IDE systems may perform better on certain data distributions than others, it is important that the attributes in the dataset exhibit different types of distributions, and contain random as well as correlated data. 
Furthermore, systems like approXimateDB \cite{li2016wander} implement online joins while classic systems such as MonetDB \cite{monetdb} typically use a blocking join such as a radix hash-join.
To be able to measure the effect of such joining techniques, an IDE benchmark must support different schema complexities, i.e., normalized as well as de-normalized star schemas.

\paragraph*{Customizability} 

We believe that in order for  an IDE benchmark to be adapted by the community, it is crucial that workloads and datasets can be customized to the use case of an IDE system.
A benchmark should therefore facilitate the ability to create workloads based on modifiable configurations, and to scale any seed datasets to an arbitrary size while preserving the original distributions.

\section{The IDEBench Design}
\label{sec:benchmark}

In this section we explain how we designed \idebench{} along the requirements for workloads, data and metrics, and describe how the results of the final metrics are computed and reported.

\subsection{Overview}
\label{overview}

IDEBench comprises three main components:
1) A \emph{data generator} that scales any seed dataset to an arbitrary size.
2) A \emph{workload generator} that create sequences of interactions, which we henceforth refer to as ``workflows''.
3) An \emph{benchmark driver}, which runs/simulates workflows, delegates interactions to  system drivers, and generates reports.
4) Different \emph{system adapters}, which are custom proxies between the database systems under test and the interpreter which runs a workload. 
The drivers are responsible to translate the benchmark workload into queries supported by a system, and returns computed results back to the interpreter.
\subsection{Data Generator}
\label{sec:data}

\begin{figure}
\centering
\includegraphics[width=0.4\columnwidth]{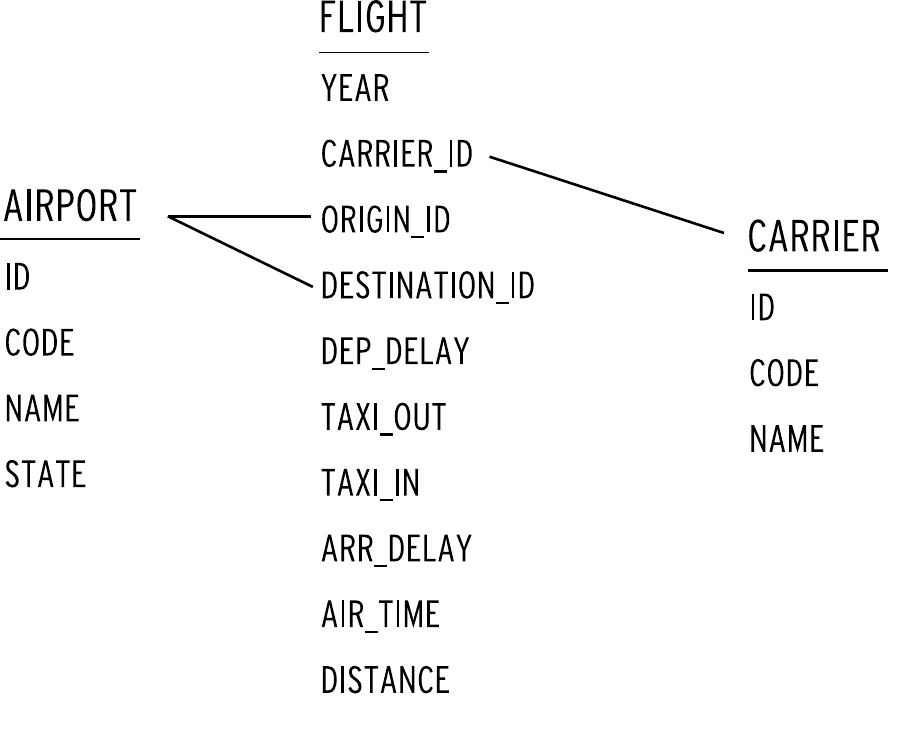}
\caption{The schema of the default dataset}
\label{fig:dataset}
\end{figure}

The default configuration of \idebench{} uses a real-world data set containing U.S. domestic flights \cite{BTS} (see Figure \ref{fig:dataset}).
We use this dataset instead of existing data generators from TPC-H or TPC-DS since it contains real-world data and distributions.
This is important because the underlying distributions can affect quality of results, especially in the case of database systems that use approximate query processing techniques.
In the default configuration, the benchmark use this dataset to test the systems. 
Alternatively, users can use any other dataset to customize the benchmark.

In order to scale the default data set (but also custom data sets), \idebench{} comes with a data generator uses a seed datasets to create a new dataset of arbitrary size.
The generator tries to maintain distributions in the data and relationships between attributes when scaling.
It also supports the transformation of data into a more normalized form (e.g., one fact and multiple dimension tables) based on a specification given by the user.
In the default configuration, the benchmark uses three data sizes (S, M, and L) for each data set to test the runtime for increasing data sets.
We elaborate more on the concrete data set sizes in the evaluation section.

The data scaling procedure to scale a data set works as follows:
From the seed dataset we first create a random sample. 
We then compute the covariance matrix $\Sigma$ and perform the Cholesky decomposition on $\Sigma = A^{T}A$.
To create a new tuple, we first generate a vector $X\sim\mathcal{N}(0,1)$ of random normal variables and induce correlation by computing $\widetilde{X} = AX$.
We then transform  $\widetilde{X}$ to uniform distribution and finally use the CDF from our sample to transform the uniform variables to a correlated tuple.
Optionally, as a last step the data generator then vertically partitions the data into multiple tables (normalization) based on a user-given schema specification.
\subsection{Workload Generator}
\label{sec:workflowtypes}

\begin{figure*}
\hspace{-3ex}
\includegraphics[width=1.05\textwidth]{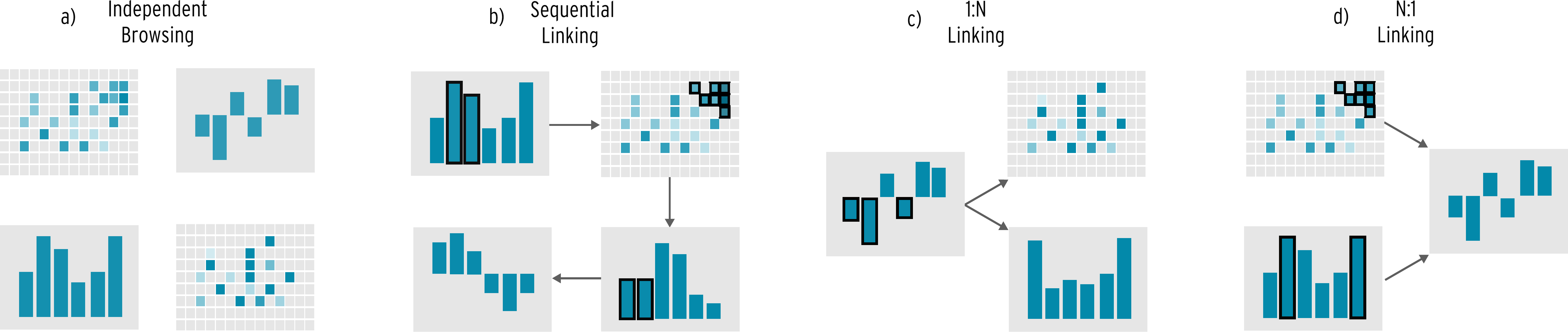}
\caption{Illustrates four different workflow types used in \idebench{} workloads. a) shows four different independent visualizations; filtering/selecting any of the visualizations does not affect the others. b,c and d) show how filtering/selecting data on a source visualizations force linked visualizations to update.}
\label{fig:workflowtypes}
\end{figure*}

Unlike the static workloads of traditional benchmarks, \idebench{} aims to measure the performance of database workloads that result from interactive data exploration frontends over the course of entire user-centered workflows (see Section \ref{sec:landscape}).
In these workloads it is common that queries are built and refined incrementally, executed with think-times between queries rather than being processed back-to-back, and that oftentimes multiple queries need to be processed simultaneously in order to update multiple linked visualizations.

\paragraph*{Workload Generator}

\begin{figure}
\centering
\includegraphics[width=0.98\columnwidth]{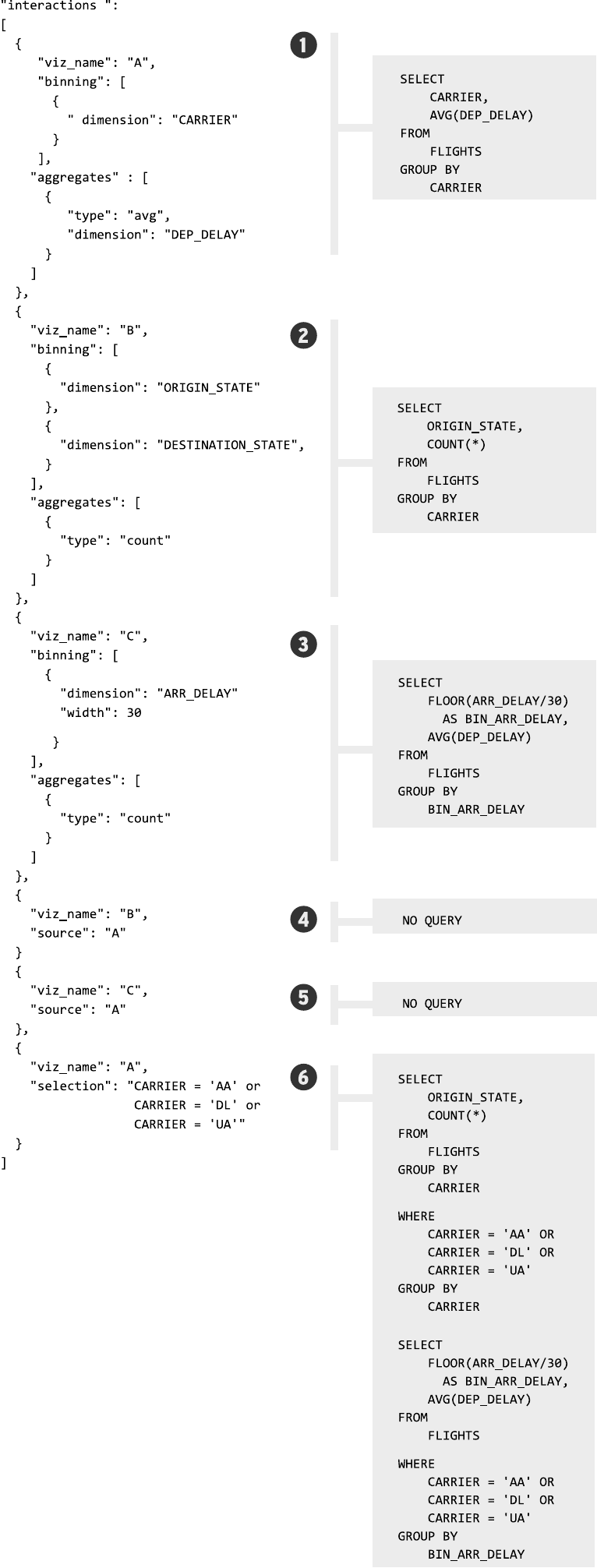}
\caption{A JSON-based specification and translation to SQL queries for the 1:N workflow in Figure \ref{fig:workflowtypes}c.}
\label{fig:workflow_json}
\end{figure}

In order to reflect this in our benchmark, \idebench{} workloads simulate user interactions that are typical interactions in IDE frontends.
We have implemented a workflow generator that allows users of \idebench{} to create custom workflows of for any of the four types below.
Workflows are sequences of common interactions that resemble IDE interaction patterns of real users (see Figure \ref{fig:workflow_json}.
The workflow generator models workflows as Markov Chains with pre-defined (and customizable) probability distributions for each of the workflow types to sample a sequence of interactions and filter/selection criteria.
We base the different types of workloads and probability distributions of various interactions on observations made by analyzing the logs and videos of past user studies \cite{idea, zgraggen2017progressive}.

\paragraph*{Workflow Types}
\begin{itemize}
    \item \textit{Independent Browsing} (Figure \ref{fig:workflowtypes}a), where users explore a dataset by creating visualizations of different dimensions of the data and by applying filters that only affect a single visualization at the time.
    This is often applied by users to get a first quick overview of the data.
    \item \textit{Sequential Linking} (Figure \ref{fig:workflowtypes}b), where users create multiple visualizations that are (logically) sequentially linked.
    Filters and selections on these visualizations trigger multiple concurrent queries to update all affected visualizations.
    This type of workflow is often used for targeted explorations where users drill down in the data to verify one concrete hypothesis.
    \item \textit{1:N Linking} (Figure \ref{fig:workflowtypes}c), where filtering/selecting on one visualization triggers $N$ other queries for directly linked visualizations.
    This type of workflow is often used to see how different subsets of the data matching certain criteria affect visualizations of other dimensions of the data.
    \item \textit{N:1 Linking} (Figure \ref{fig:workflowtypes}d), where applying a filter to any one of $N$ visualization affects a single directly linked visualization.
    This is often used to incrementally build filter expressions involving multiple dimensions.
\end{itemize}

These patterns determine the workload for the database system regarding the fact of how many queries are triggered simultaneously.
For example, for the independent browsing type one user interaction results on in one query since only one visualization needs to be updated.
In contrast, for the 1:N linking, one user interaction (e.g., changing the filter) on a visualization early in the sequence can trigger many queries since multiple visualization potentially need to be updated.

Each generated workflow comprises a sequence of interactions performed by users: Creating a visualization i.e., formulating and executing query (interaction 1, 3, 4 in Figure \ref{fig:workflowtypes}), filtering/selecting (interaction 2 and 6), linking visualizations (interaction 5), and discarding a visualization (not shown).
Alternatively, customized workflows can additionally be built manually to match a specific usage scenario. However, they are not part of the default configuration of \idebench{}.
Once generated, they can be inspected with an interactive viewer.

\subsection{Benchmark Driver}
The core of \idebench{} is an benchmark driver, a simple command line application (written in Python) configured to load and simulate workflows by forwarding interactions to system adapters (see Section \ref{drivers}), and to evaluate the results that are fed back into the driver.
When running a workflow the driver keeps track of a visualization graph (similar to \ref{fig:workflow_json}), notifying an adapter about which interactions have been executed.
These interactions are specified in a JSON-based format (see Figure \ref{fig:workflow_json}).
The benchmark driver automatically translates queries to SQL, or alternatively, lets the system driver translate queries into a language compatible with the system being evaluated.

\subsection{System Adapters}
\label{drivers}
To be evaluated by our benchmark a system needs to implement a driver interface that acts as proxy between the benchmark and the system under test. 
The benchmark driver delegates interactions from the workflows to the system adapter.
The system adapter takes these interactions and translates them into queries executable by the system under evaluation (e.g., SQL in case of a classical analytical database system). 
However, any other query language or proprietary API can be used by implementing an additional adapter.
\begin{lstlisting}[language=python, basicstyle=\tiny\ttfamily, caption={A stub for a system adapter.}]

class SampleAdapter:
    
    def process_request(self, viz_specification):
        # 1. translate to a query format understood by the system
        # 2. execute queries
        # 3. fetch and parse result
        # 4. write results back to interpreter
        
    def link_vizs(self, viz_from, viz_to):
        # use the logical links as hint for speculative
        # query execution, if applicable
        
    def delete_vizs(self, vizs):
        # free memory, if applicable
    
    def workflow_start(self):
        # called before a workflow starts
        
    def workflow_end(self):
        # called when a workflow is done
\end{lstlisting}

\subsection{Settings}
\label{sec:settings}
In the following, we show the most important parameters that our benchmark uses to test different configurations.

\begin{center}
\small
\begin{tabular}{| p{2cm} | p{5.5cm} |}\hline
    Time Requirement (TR) & The maximum execution duration for a query. \\ \hline
    Dataset and Size & The dataset to run the benchmark on and the number of tuples to up- or downsample dataset to. \\ \hline 
    Think Time & The delay between two consecutive interactions.  \\ \hline
    Using Joins & Whether a normalized or de-normalized schema is used. \\ \hline
    Confidence Level & The confidence level at which an AQP returns margins of error. \\\hline
\end{tabular}
\end{center}

\idebench{} provides default configurations for these parameters but we also allow users of our benchmark to vary the parameters such that they can test a system with settings that match the requirements of their application.
However, in order to enable comparability of benchmark results the default configurations should be used.
The details of the default configurations are listed in the evaluation section.

\vspace{0.3cm}

\paragraph*{Time Requirement}

Various researchers have identified the interaction response time as a crucial factor in interactive data exploration systems ~\cite{heer2012interactive,hanrahan2012analytic,zgraggen2017progressive}.
Liu et. al.~\cite{liu2014effects}, for instance, showed that even response times of about $500$ms could lead to poor user performance.
Studies of other tasks argue for even lower thresholds ~\cite{brutlag2009speed, beigbeder2004effects}.
In general, it is important that the speed of the systems aligns with the users' speed of interaction.
Acceptable response times may vary depending on the domain and task.
For example \cite{seow2008designing, nielsen2009powers, card1991information, shneiderman1984response} claim that responses within one second allow users to stay focused, while response times over ten seconds exceed the average attention span.
In our experiments, we used time requirements of $0.5$s, $1$s, $3$s, $5$s and $10$s.

\paragraph*{Dataset and Size} This parameter represents the dataset to be used in the benchmark as well as its size in de-normalized form (the size of the fact table).
Normalization of the data is applied as a post processing step after generation.

\paragraph*{Think Time}
User studies have shown that there are significant delays (think time) between two consecutive interactions \cite{idea}, which systems can leverage to run speculative queries.
For the benchmark we recommend values between 3 and 10 seconds.

\paragraph*{Using Joins}
While many systems only support queries on data in de-normalized form, e.g. \cite{idea, dice}, \idebench{} can also be run to measure query performance on more normalized datasets (e.g., represented as a star schema) .

\paragraph*{Confidence Level}
Many approximate query processing (AQP) systems are capable of returning confidence interval for approximate results, and let users define the desired confidence level.
\idebench{} uses $95$\% as default.
\subsection{Metrics}
\label{sec:metrics}

As discussed before, the metrics of our benchmark should reflect the interactivity and the quality of the results throughout all operations of the executed workflows.
For each executed query we evaluate the following metrics and aggregate them into a final report (see next section):

\paragraph*{Time Requirement Violated}
Time Requirement (TR) Violated is a boolean value indicating whether or not a query violated the time requirement specified in the settings (see Section \ref{sec:settings}).
TR is violated if time TR after initiating the query, no result is present or can be fetched.
In practice this means, that for batch-processing and APQ systems, TR is violated if the run-time of a query is greater than TR, and no intermediate result is present.
For progressive system, TR is violated if time TR after initiating a query no result can be fetched.
\idebench{} measures a boolean rather than the actual duration as time violation in order to guarantee constant run-times for any workflow; queries whose run-time exceed TR are cancelled.

\begin{center}
\small
\begin{tabular}{| p{2cm} | p{5.5cm} |}\hline
    Time Requirement (TR) Violated & Boolean whether a query violated the time requirement Time Requirement (TR). \\ \hline
    Missing Bins & The ratio of the number of bins bins for which no result has been delivered and the total number of bins in the ground-truth. \\ \hline
    Mean Relative Error & The mean relative error of all bins returned in the result (see definition below). \\ \hline
    Cosine Distance & A measure of how much the ``shape'' of a result resembles the ground-truth. \\ \hline
    Mean Margin of Error & The mean of all relative margins of error for all bins. \\ \hline
    Out of Margin & The number of approximate results that were outside of the return confidence interval.  \\ \hline
    Bias & The sum of all returned values in a result divided by the sum of all true results for the bins returned. \\ \hline
\end{tabular}
\end{center}

\paragraph*{Missing Bins/Groups} Missing Bins/Groups 
is the ratio of all bins for which no result has been delivered, and all bins in the ground-truth.
It is a measure of completeness for an aggregate query result, irrespective of the number of tuples processed by a system. 

\[ \text{Missing Bins} = \frac{|bins\_missing|}{|bins\_in\_groundtruth|}\]

\paragraph*{Mean Relative Error}
To measure the error between a result of an aggregate query and its ground-truth we compute the relative error; i.e., the ratio between the difference on the estimated result $F_i$ and the actual result $A_i$.

\[ \text{Mean Relative Error} = \frac{1}{n} \sum\limits_{i=1}^{n} \frac{|F_i - A_i|}{|A_i|} \]

We use the mean relative error due to its popularity in existing literature and ease of interpretation.
However, it is important to note that the relative error is not defined for any $A_i = 0$, which is especially problematic for aggregate functions such as \texttt{AVG}, \texttt{MIN/MAX}, or \texttt{SUM}, if the expected value is zero.
A possible future alternative is the Symmetric Mean Absolute Percentage Error, which is defined as:
\[ \text{SMAPE}  = \frac{1}{n} \sum\limits_{i=1}^{n} \frac{|F_i - A_i|}{|F_i| + |A_i|} \]

While less intuitive, SMAPE is defined for $A_i = 0$, unless $|F_i| = 0$, in which case the error is $0$.
SMAPE is also bounded at 0 and 1, which may simplify interpretation across different experiments.

\paragraph*{Cosine Distance} In some cases users may be more interested in the relative difference of aggregated results, i.e. the distribution of values, rather than the true values.
We measure the cosine distance to test how much the ``shape'' of a result deviates from its ground-truth.
For instance, it captures if a system is able to provide a good estimate of the relative frequency distribution of the data, even if the relative errors are high.
To make sure both result vectors $F$ and $A$ are of equal length, we set the value at each missing bin to zero.

\[\text{Cosine Distance} = 1 - \frac{\sum\limits_{i=1}^n F_i A_i}{\sqrt{\sum\limits_{i=1}^n F_i^2}\sqrt{\sum\limits_{i=1}^n A_i^2}}  \]

\paragraph*{Mean Margin of Error} Approximate and progressive query system typically provide confidence intervals with their query results.
To get a sense of how tight these intervals are, i.e. how likely the returned result was just a good guess, we compute the mean and standard deviation of all relative margins of error.

\paragraph*{Out of Margin} Out of Margin is a sanity check to test whether the system returns results roughly at the confidence level specified in the settings.
We measure the number of how many of the per-bin results exceeded the returned margins of error.

\paragraph*{Bias} Indicates whether a system tends to over or under-estimate aggregated values.
This metric becomes especially important if other error metrics are employed, as some (e.g., SMAPE) penalize under/over-estimation unequally.
\vspace{6ex}
\subsection{Reporting}

\begin{figure*}[ht]
\centering
\hspace{-3ex}
\includegraphics[width=\textwidth]{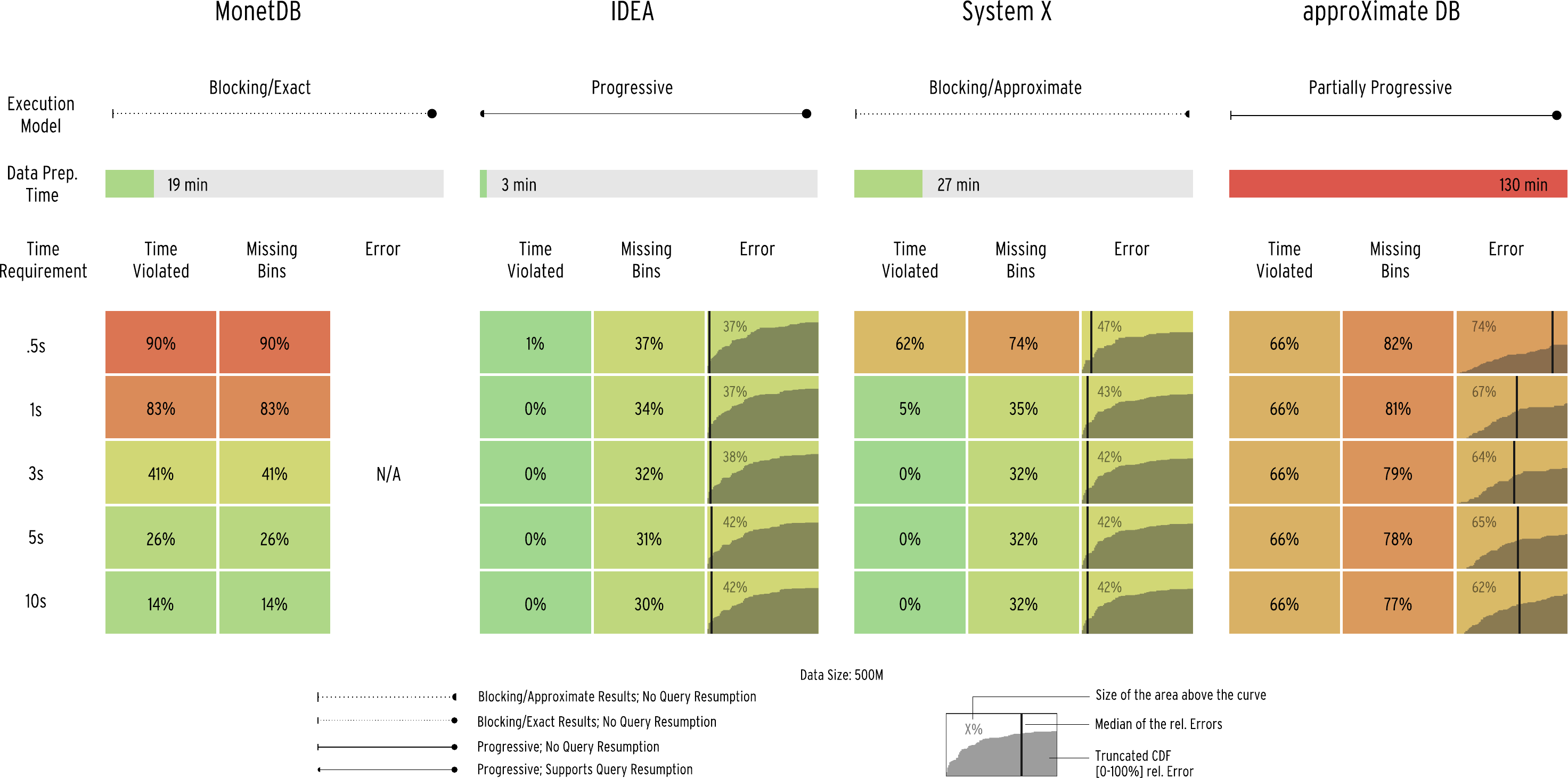}
\caption{Shows the aggregated benchmark results for four systems in a summary report. The benchmark was run for five time requirements on a fixed dataset of size 500M. It shows the mean percentage of time violations and missing bins, as well as a visualization of the behavior of the mean relative errors (MREs). It shows a CDF of the MREs, truncated to errors less or equal to 100\%. Thus, the greater the proportion of small errors, the smaller the area above the curve (shown as percentage above the CDF).}
\label{fig:overall_results}
\end{figure*}

Upon completion of running the benchmark, \idebench{} generates two reports:
1) An aggregated summary report listing how frequently the time requirement was violated, how many bins are missing on average, and the distribution of mean relative errors for all queries which did not violated the time requirement.
Figure \ref{fig:overall_results} shows an example of such a summary report.
2) A detailed report listing all settings and metrics on a per query basis.

Ideally, a user wanting to explore a new dataset can effortlessly plug in the data.
Therefore, users of \idebench{} are required to report on all actions needed to be taken to prepare for a benchmark run (called ``data preparation time'' in our report); i.e., the time from connecting to a new data source to being actually able to start running the workload.
This includes steps and time taken to copy the dataset into the system to create sample tables/views offline, perform pre-processing, warm-up queries, etc.
\vspace{6ex}

\section{Evaluation}
\label{sec:results}

To demonstrate the applicability of \idebench{} across various systems types, we executed the benchmark on the following systems:
(1) \emph{MonetDB:} A state-of-the-art open-source analytical column-store DBMS, which uses a blocking query execution model that requires users to wait until an exact query result is computed.
Thus, upon initiating a query, the run-time of the query is unknown.
(2) \emph{approXimateDB/XDB:} A PostgreSQL-based DBMS that supports online aggregation using the wander join algorithm \cite{li2016wander}.
It allows for a maximum run-time to be set when initiating a query.
It additionally supports a ``report interval'', so that intermediate results can be retrieved at fixed time intervals.
XDB has some limitations in terms of query support, which we describe in detail in Section \ref{sec:results:exp1}. 
(3) \emph{IDEA:} A system that supports online aggregation and has a fully progressive computation model where, after initiating a query, results can be polled at any point in time.
(4)  \emph{System X:} A commercial in-memory AQP system that operates on stratified sample tables (offline sampling).
The run time of queries cannot be set explicitly, but must be specified by means of setting the size of samples tables, i.e. the sampling rate.
(5) \emph{System Y:} A commercial specialized engine for IDE, which provides an in-memory optimization layer on top of a number of DBMS systems.

In the remainder of this section, we first describe the general setup of all our default configurations (Section \ref{sec:results:setup}), report on the data preparation time of each of the systems described above (Section \ref{sec:results:coldstart})  and present the overall benchmark performance (Section \ref{sec:results:exp1})  as well as experiments with benchmark settings (Section \ref{sec:results:exp2} to Section \ref{sec:results:exp6}).

\vspace{3ex}
\subsection{Default Configurations and Setup}
\label{sec:results:setup}

\subsubsection*{Default Configurations} In the following, we discuss the default configurations that our benchmark defines. 
In the default configuration, \idebench{} uses the flight dataset (see Section \ref{sec:data}) with S=100 million, M=500 million, and L=1 billion tuples in the de-normalized form (i.e., only one large table with all attributes).
Moreover, the default configuration runs $10$ workflows for each of the workflow types described in Section \ref{sec:workflowtypes}, as well as $10$ ``mixed'' workflows which exhibit usage patterns from all four workflow types.
As parameters in the default configuration, we use five different time requirements 0.5s, 1s, 3s, 5s and 10s, with a confidence level set to 95\%.
While most recommended latencies are in the range of 0-1s, we also included greater ones in the default configuration to get a better understanding of how fast results converge.
Finally, we use ten different think-times ranging from 1s to 10s (see Figure \ref{fig:exp_latency}) in our default configuration.
We empirically found these to be good estimates by analyzing video logs of a previous user study \cite{vizdom}.

\subsubsection*{Setup} We ran the benchmark on \emph{MonetDB}, \emph{approXimateDB}, \emph{IDEA} and \emph{System X}.
In order to stress-test the systems, we only report for the think-time of $1s$ in all experiments except experiment 3 (see Section \ref{sec:results:exp4}), in which we analyzed the effect of varying the think-time.
We did not run the full benchmark on \emph{System Y} since it does not have a publicly available API.
For \emph{System Y} we executed selected workflows manually over its user interface in a separate experiment (see Section \ref{sec:results:exp6}).
All experiments were conducted on a computer with two Intel E5-2660 CPUs (2.2GHz, 10 cores, 25MB cache) and 256GB RAM.
We use the default configuration for all systems, and abstain from optimizations for any of the system parameters.

\subsection{Exp. 1: Overall Results}
\label{sec:results:exp1}

\begin{figure*}[t]
\centering
\hspace{-3ex}
\includegraphics[width=\textwidth]{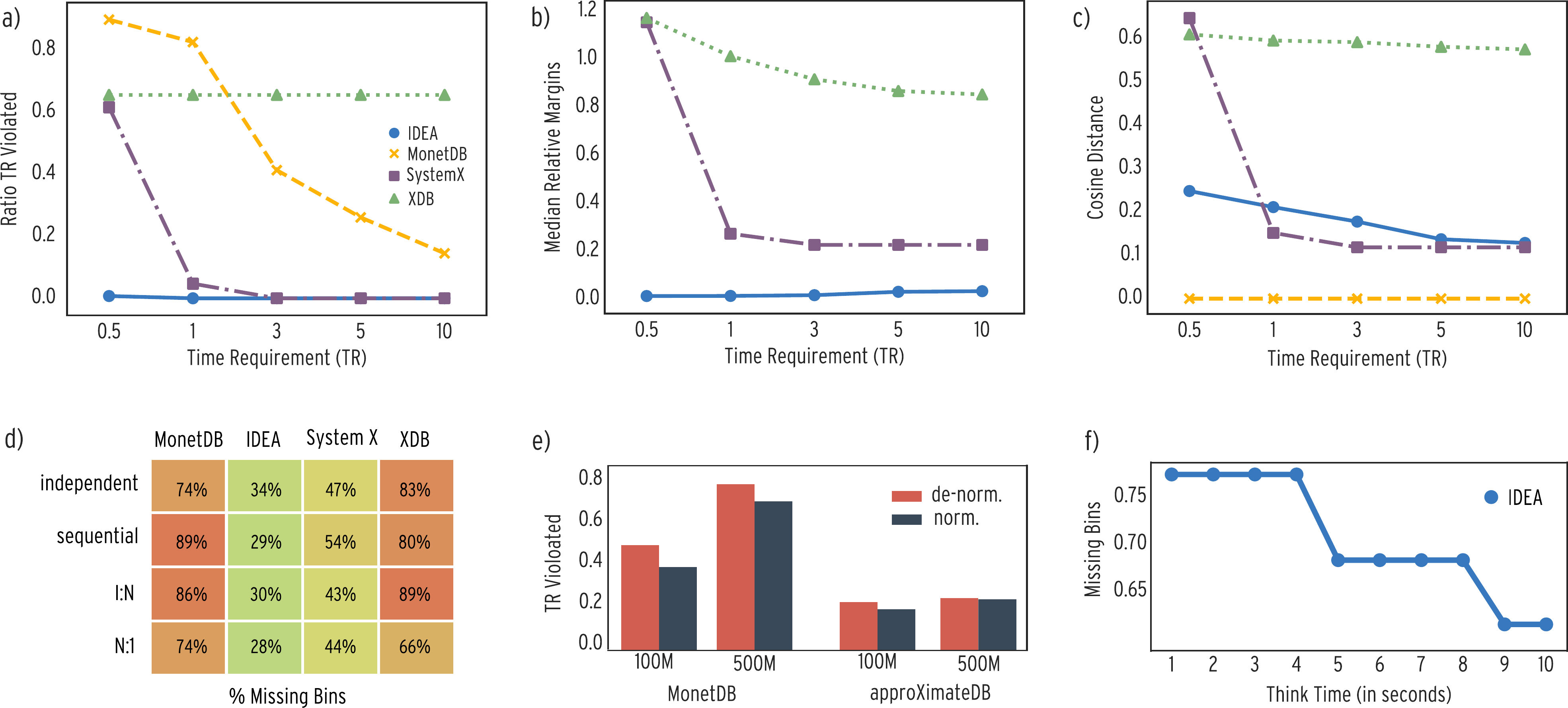}
\caption{
a, b, and c) Show how the ratio of TR violations, the median of the mean relative margins, and the cosine distance develop with increasing time requirements.
d) Compares how the proportion of missing bins differs based on which system and workflow type is  used.
e) A comparison of the proportion of violated time requirements for MonetDB and \emph{approXimateDB}, using a normalized and de-normalized dataset of size 500M.
f) Shows the effect of varying think-times on missing bins.}
\label{fig:exp_latency}
\end{figure*}

In our main experiment (Figure \ref{fig:overall_results} and \ref{fig:exp_latency}) we analyzed how the four systems behave with respect to different time requirements (see Figure \ref{fig:overall_results}).
We show the results the mixed workload, i.e., 10 workflows with a mix of the four exploration behavior described in Section \ref{sec:workflowtypes}, data size of 500M, a de-normalized schema, and a confidence level to 95\%.

\subsubsection*{Data Preparation Time}
\label{sec:results:coldstart}
In \emph{MonetDB} data stored in a CSV file can be loaded into the database through an SQL interface, which takes 19 minutes for 500M records.
There is no pre-processing time upon starting the server.
\emph{approXimateDB} works identically, but takes 130min (time split between adding the data and adding a primary key).
This system also provides support for an additional SQL statement to pre-load relations and indexes into the database buffer in main memory.
For our experiments we did not make use of this statement.
\emph{IDEA} expects data in a single CSV file and does not need any pre-processing.
On start-up, \emph{IDEA} by default loads a fixed amount of tuples into main memory, which takes 3min.
In \emph{System X} data stored in a CSV file can be loaded into the database through a SQL interface.
In order to be able to execute approximate queries, stratified sample tables have to be created offline.
We used a sample size of 1\% of the data size.
\emph{System X} further requires upon restart of the systems that each connection must execute a warm-up query.
For 500M records we measured a data preperation time of 27min.

\subsubsection*{Speed and Quality Metrics}
As expected for an exact execution model, \emph{MonetDB}'s TR (Time Requirement) violations decrease roughly linearly with the defined TR, and so does the proportion of missing bins (see also Figure \ref{fig:exp_latency}a).
\emph{approXimateDB}, on the other hand, violates the time requirement consistently around ~66\% for any TR.
This is due to the fact that while \emph{approXimateDB} supports online aggregation for \texttt{COUNT} and \texttt{SUM}, it does not provide online support for \texttt{AVG} nor for multiple aggregates in a single query.
Thus, queries that cannot be executed online typically fail for a small TR.
We therefore set up \emph{approXimateDB} so that any query that cannot be executed online will fall back to a regular Postgres query.
With \emph{System X} more than 50\% of all queries violate TR=0.5s.
Interestingly, though, for TR=1s only 5\% are violated, and for TR=3s all query results are returned on time.
The percentage where TR is violated is therefore a good indicator of how large one should set the sample size, if speed is more important than result quality.
\emph{IDEA} on other hand does not violate any TR, with the exception of 1\% of all queries for TR=0.5.
The authors confirmed that this is due a slightly higher overhead for the first query after a restart of the system.
\emph{IDEA} also starts off with significantly less missing bins (37\%) for TR=0.5 than any other system, but achieves similar values to \emph{System X} for TR>=1s.
Furthermore, \emph{IDEA} manages to perform better than other systems in terms of mean relative error of all return results.
The median of all mean relative is constantly much lower than \emph{approXimateDB}'s, and marginally lower than the one of \emph{System X}.
More interestingly, \emph{approXimateDB}'s area above the curve is much higher than the one of \emph{IDEA} and \emph{System X}, indicating that high mean relative errors occur more frequently.
A similar conclusion can be drawn by looking at the end of the curve.
\emph{approXimateDB}'s curve ends, below 50\%, indicating that more than 50\% of all mean relative error are greater than 100\%.

Figure \ref{fig:exp_latency}b and \ref{fig:exp_latency}c show how the median of the mean relative margins, and the cosine distance develops with increasing time requirements.
\emph{approXimateDB} has significantly higher relative margins than both \emph{IDEA} and \emph{System X}.
However, while \emph{System X}'s median is close to 120\% for TR=0.5s and drops to slightly above 20\% for TR=1s, \emph{IDEA}'s median remains constant around zero for all TRs.
Figure \ref{fig:exp_latency}d compares how the proportion of missing bins differs based on which system and workflow type is used.
As none of the systems we used in the evaluation use speculative execution by default, there are only few significant differences.
For instance, \emph{MonetDB} has fewer missing bins on average for independent browser and N:1 workflows, which may be attributed to the fact that any interaction of these workflows only trigger a single query.

\subsection{Exp. 2: Normalized vs. De-normalized}
\label{sec:results:exp2}

In a third experiment, we compare the performance of \emph{MonetDB} and \emph{approXimateDB} using a normalized and de-normalized schema. 
We exclude \emph{IDEA} as it does not support joins. 
Similarly, \emph{System X}, only works on de-normalized data.
We used our data generator to create two datasets of 100 million and 500 million tuples and normalized the data so that the fact table holds foreign keys to two dimension tables (airports and carriers).
Interestingly, as can be seen in Figure \ref{fig:exp_latency}e, both \emph{MonetDB} and \emph{approXimateDB} perform slightly better in terms of time requirement violations with a normalized schema.
We can attribute this to the fact that the overall size of all tables is reduced since splitting data into fact and dimension tables reduced the overall database size.
Another interesting observation is that \emph{MonetDB}'s proportion of TR violations grows with the size of the normalized dataset.
Conversely, \emph{approXimateDB} is able to keep it roughly at the same level, due to its online join support for aggregate queries.

\subsection{Exp. 3: Varying Think-Time}
\label{sec:results:exp4}
In this experiment, we evaluated the impact of increasing think times between interactions (see Figure \ref{fig:exp_latency}f).
We used an experimental extension of \emph{IDEA} that speculatively executes queries when two visualizations are linked.
For the setup we used a fixed data size of 500M tuples, a time requirement of 3s, and created a custom workflow comprising following four interactions.
1) data for a 2D count histogram (100 bins) of arrival delays vs. departure delays is requested.
2) data for a 1D count histogram (25 bins) of carriers is requested.
3) a link between the to visualizations is established, setting the 1D histogram as source and the 2D histogram as target
4) a single carrier is selected in the 1D histogram, forcing the 2D histogram to update

\emph{IDEA} keeps track of a visualization graph, and executes queries for every possible single bin selection in the source visualization.
If upon the next interaction one of the bins in the 1D histogram is selected, \emph{IDEA} can return a potentially better estimate of the results, as the query has had more processing time.
Figure \ref{fig:exp_latency}f shows the proportion of missing bins across ten think times (1s - 10s).

\subsection{Exp. 4: Other Effects}
\label{sec:results:exp5}
We also used the benchmark results to analyze each system for other effects, e.g., differences in performance in regards to bin widths/number of bins, binning types (1D vs .2D, nominal vs. quantitative ranges), as well as how systems respond to interactions that lead to multiple concurrently executed queries.
We analyzed all queries listed in the detailed reports of all systems, but found no evidence that any of the factors above have a significant impact on the performance of any of the metrics.
By far the most crucial factor in terms of query performance, seems to be the specificity of filter/selection predicates.

\subsection{Exp. 5: Experiment with System Y}
\label{sec:results:exp6}

In the last experiment we replicated a selected subset of our workflows in a commercial IDE System Y and used MonetDB as a backend.
We used a fixed data size of 500M and simulated three variants of the 1:N workflow type.
In particular, we were interested to see if \emph{System X} uses an intermediate layer that pre-fetches/computes results, similar to the experimental extension of \emph{IDEA} (see Section \ref{sec:results:exp5}).
However, we did not find this to be the case.
System Y renders and updates the visualizations in the workload roughly at the same speed as when one uses \emph{MonetDB} directly, with an added delay of about 1-2s per query.
This is likely to be the rendering overhead to draw the visualizations.

\section{Discussion and Future Work}
\label{sec:discussion}

\paragraph*{Main Findings}
By implementing \idebench{} for four systems we have shown that the performance in terms of data preperation time, TR violations as well as quality of the results can vary significantly.
We saw that progressive and AQP systems like \emph{IDEA} and \emph{System X} were able to keep time violations at a minimum while maintaining low error rates with increasing data sizes and time requirements.
This is in stark contrast to classical analytical databases represented by \emph{MonetDB} where time violations increase for larger datasets and time requirements.
We saw that \emph{approXimateDB} can only execute a subset of the queries in our workload online.
It has to revert to executing other queries in a blocking fashion, which in turn leads to significantly more TR violations.
For AQP systems where sample tables need to be created offline (e.g., \emph{System X}),  quality of result metrics such as the relative error and missing bins remain constant across different time requirements.
Thus, such a system would have to scan the full table or to create additional sample tables of different sizes in order to achieve a higher result quality.
This in turn would increase the data preparation time.
Furthermore, our results indicate at which point the use of an AQP that implements online sampling over offline sampling is beneficial; stratified sampling is able to provide results similar to online systems. However, there is significant overhead for offline sample-based approaches.
Determining a ``good'' sample size to find a good trade-off between speed and quality of the results is time-consuming, and sample tables created offline cannot be fully optimized because in IDE the workload is unknown ahead of time.

\paragraph*{Future Work}
The core idea of \idebench{} is to provide a benchmark that simulates typical user behavior for basic tasks in IDE such as such as executing aggregate queries, filtering result sets, etc. \cite{keim,visscientific}.
The richer the tasks the harder they become to benchmark. 
For the future we therefore envision an extensible benchmark design that defines different task specific core-sets where each core-set aims to analyze a different functional aspects; e.g., one core-set (as defined in this paper) only tests simple analytical operations, whereas another one tests more complex model building tasks.
Which core-sets are used to evaluate a system depends on the supported functionality of that system.

Specifically, we envision for our benchmark four core-sets:
{\em Core-Set I} focuses on Interactive Data Exploration as covered by the benchmark implementation presented in this paper. However, it excludes interactive model building, which is part of {\em Core-Set II}. 
For the future version of \idebench{}, we believe that {\em Core-Set I} and {\em II} should be coupled as it seems unreasonable to assume that one would build a model without having the possibility to inspect the data set before.
{\em Core-Set III} should then concerned with benchmarking  recommendation engines which are often used to complement IDE systems by steering users to interesting parts of a new data set. Examples systems are SeeDB \cite{seedb} or Data Polygamy \cite{polygamy}.
Finally, for {\em Core-Set IV} we suggest to extend the benchmark to other data sets that allows us compare the interactive data cleaning capabilities covered by systems such as DataWrangler \cite{datawrangler}, Trifacta \cite{trifacta} or Paxata \cite{paxata}).

We believe that there will be variety of systems that can only cover the functionality of some the core-sets above.
We therefore envision that each core-set can be tested individually.
The higher the core-set number the harder it is to define a benchmark since the sheer complexity of supported operations is increasing and their comparability becomes more difficult.

Finally, a last important aspect is to make the benchmarking code publicly available on the web \cite{idebench} and include more recent benchmarking results on the already tested systems as well as results on other systems.
Moreover, we plan to allow other research groups as well as industry to upload other data sets and user-defined workflows in the format that they can be included our framework to cover interesting aspects of other domains.
\section{Conclusion}
\label{sec:conclusion}

In this paper, we presented a new benchmark \idebench{} for evaluating systems for interactive data exploration (IDE).
\idebench{} defines a new set of metrics and a workload generator to simulate different exploration behaviours of users as well as a data generator to better address the challenges of IDE workloads.
Based on this new benchmark, we conducted an evaluation study that covered five different systems (approximateDB, IDEA, MonetDB as well as two commercial systems) of three different categories (traditional analytical database systems, approximate query processing engines, as well as specialized engines for IDE) that are used today in order to execute IDE workloads.

\begin{appendices}
\section{Appendix}

\subsection{A Detailed Benchmark Report}
\label{appendix:report}
\begin{table*}
\scriptsize
\centering
\caption{An example of a detailed benchmark report for a single workflow.}
\label{tab:detailed_report}
\resizebox{\textwidth}{!}{
\begin{tabular}{lllllllllllllllllllllll}
\hline
id & interaction & viz\_name & driver & data\_size & think\_time & time\_req & workflow & start\_time   & end\_time     & tr\_violated & bin\_dims & binning\_type              & agg\_type & bins\_ofm & bins\_delivered & bins\_in\_gt & rel\_error\_avg & rel\_error\_stdev & missing\_bins & cosine\_distance & margin\_avg & margin\_stdev \\ \hline
0  & 0           & viz\_0    & idea   & 500m       & 3000        & 500       & mixed\_2 & 1514643736024 & 1514643736547 & FALSE        & 1         & quantitative               & avg       & 8         & 38              & 56           & 0.02            & 0.02              & 0.32          & 0.18             & 0.05        & 0.06          \\
1  & 1           & viz\_1    & idea   & 500m       & 3000        & 500       & mixed\_2 & 1514643736566 & 1514643737107 & FALSE        & 1         & quantitative               & count     & 1         & 82              & 159          & 1.39            & 3.47              & 0.48          & 0.00             & 0.99        & 0.82          \\
2  & 2           & viz\_2    & idea   & 500m       & 3000        & 500       & mixed\_2 & 1514643737128 & 1514643737663 & FALSE        & 1         & nominal                    & count     & 6         & 53              & 53           & 0.01            & 0.02              & 0.00          & 0.00             & 0.01        & 0.03          \\
3  & 3           & viz\_3    & idea   & 500m       & 3000        & 500       & mixed\_2 & 1514643737686 & 1514643738381 & TRUE         & 2         & quantitative\_quantitative & avg       & 87        & 507             & 1244         & 0.09            & 0.17              & 0.59          & 0.39             & 0.15        & 0.19          \\
4  & 4           & viz\_2    & idea   & 500m       & 3000        & 500       & mixed\_2 & 1514643738414 & 1514643738943 & FALSE        & 1         & nominal                    & count     & 6         & 53              & 53           & 0.01            & 0.02              & 0.00          & 0.00             & 0.01        & 0.03          \\
5  & 5           & viz\_2    & idea   & 500m       & 3000        & 500       & mixed\_2 & 1514643738964 & 1514643739488 & FALSE        & 1         & nominal                    & count     & 2         & 43              & 51           & 0.49            & 0.63              & 0.16          & 0.02             & 1.05        & 0.60          \\
6  & 6           & viz\_0    & idea   & 500m       & 3000        & 500       & mixed\_2 & 1514643739508 & 1514643741705 & TRUE         & 1         & quantitative               & avg       & 9         & 22              & 56           & 0.03            & 0.03              & 0.61          & 0.38             & 0.01        & 0.02          \\
7  & 7           & viz\_2    & idea   & 500m       & 3000        & 500       & mixed\_2 & 1514643741725 & 1514643746601 & TRUE         & 1         & nominal                    & count     & 1         & 53              & 53           & 0.01            & 0.03              & 0.00          & 0.00             & 0.03        & 0.06          \\
8  & 7           & viz\_0    & idea   & 500m       & 3000        & 500       & mixed\_2 & 1514643741726 & 1514643742247 & FALSE        & 1         & quantitative               & avg       & 2         & 8               & 8            & 0.00            & 0.00              & 0.00          & 0.00             & 0.00        & 0.01          \\
9  & 8           & viz\_0    & idea   & 500m       & 3000        & 500       & mixed\_2 & 1514643746621 & 1514643747138 & FALSE        & 1         & quantitative               & avg       & 0         & 8               & 8            & 0.01            & 0.01              & 0.00          & 0.00             & 0.01        & 0.01          \\
10 & 9           & viz\_2    & idea   & 500m       & 3000        & 500       & mixed\_2 & 1514643747158 & 1514643747683 & FALSE        & 1         & nominal                    & count     & 2         & 52              & 53           & 0.06            & 0.15              & 0.02          & 0.00             & 0.14        & 0.20          \\
11 & 9           & viz\_0    & idea   & 500m       & 3000        & 500       & mixed\_2 & 1514643747159 & 1514643747674 & FALSE        & 1         & quantitative               & avg       & 0         & 1               & 1            & 0.00            & 0.00              & 0.00          & 0.00             & 0.00        & 0.00          \\
12 & 10          & viz\_0    & idea   & 500m       & 3000        & 500       & mixed\_2 & 1514643747703 & 1514643748219 & FALSE        & 1         & quantitative               & avg       & 0         & 1               & 1            & 0.00            & 0.00              & 0.00          & 0.00             & 0.00        & 0.00          \\
13 & 11          & viz\_4    & idea   & 500m       & 3000        & 500       & mixed\_2 & 1514643748239 & 1514643748771 & FALSE        & 1         & quantitative               & avg       & 30        & 79              & 179          & 0.03            & 0.05              & 0.56          & 0.34             & 0.03        & 0.05          \\
14 & 12          & viz\_4    & idea   & 500m       & 3000        & 500       & mixed\_2 & 1514643748794 & 1514643749344 & FALSE        & 1         & quantitative               & avg       & 30        & 79              & 179          & 0.03            & 0.05              & 0.56          & 0.34             & 0.03        & 0.05          \\
15 & 13          & viz\_5    & idea   & 500m       & 3000        & 500       & mixed\_2 & 1514643749368 & 1514643749891 & FALSE        & 1         & quantitative               & count     & 2         & 40              & 58           & 0.24            & 0.64              & 0.31          & 0.00             & 0.40        & 0.55          \\
16 & 14          & viz\_0    & idea   & 500m       & 3000        & 500       & mixed\_2 & 1514643749913 & 1514643750431 & FALSE        & 1         & quantitative               & avg       & 0         & 1               & 1            & 0.00            & 0.00              & 0.00          & 0.00             & 0.00        & 0.00          \\
17 & 14          & viz\_4    & idea   & 500m       & 3000        & 500       & mixed\_2 & 1514643749914 & 1514643750432 & FALSE        & 1         & quantitative               & avg       & 1         & 2               & 2            & 0.00            & 0.00              & 0.00          & 0.00             & 0.00        & 0.00          \\
18 & 15          & viz\_0    & idea   & 500m       & 3000        & 500       & mixed\_2 & 1514643750464 & 1514643750981 & FALSE        & 1         & quantitative               & avg       & 0         & 1               & 1            & 0.00            & 0.00              & 0.00          & 0.00             & 0.00        & 0.00          \\
19 & 15          & viz\_4    & idea   & 500m       & 3000        & 500       & mixed\_2 & 1514643750464 & 1514643750981 & FALSE        & 1         & quantitative               & avg       & 0         & 2               & 2            & 0.00            & 0.00              & 0.00          & 0.00             & 0.00        & 0.00          \\
20 & 16          & viz\_0    & idea   & 500m       & 3000        & 500       & mixed\_2 & 1514643751002 & 1514643751518 & FALSE        & 1         & quantitative               & avg       & 0         & 1               & 1            & 0.00            & 0.00              & 0.00          & 0.00             & 0.00        & 0.00          \\
21 & 16          & viz\_4    & idea   & 500m       & 3000        & 500       & mixed\_2 & 1514643751004 & 1514643751518 & FALSE        & 1         & quantitative               & avg       & 0         & 2               & 2            & 0.00            & 0.00              & 0.00          & 0.00             & 0.00        & 0.00          \\
22 & 17          & viz\_5    & idea   & 500m       & 3000        & 500       & mixed\_2 & 1514643751538 & 1514643752060 & FALSE        & 1         & quantitative               & count     & 2         & 40              & 58           & 0.24            & 0.64              & 0.31          & 0.00             & 0.40        & 0.55          \\
23 & 18          & viz\_2    & idea   & 500m       & 3000        & 500       & mixed\_2 & 1514643752081 & 1514643752615 & FALSE        & 1         & nominal                    & count     & 0         & 51              & 53           & 0.14            & 0.21              & 0.04          & 0.00             & 0.40        & 0.29          \\
24 & 18          & viz\_0    & idea   & 500m       & 3000        & 500       & mixed\_2 & 1514643752083 & 1514643752606 & FALSE        & 1         & quantitative               & avg       & 0         & 1               & 1            & 0.00            & 0.00              & 0.00          & 0.00             & 0.00        & 0.00          \\
25 & 18          & viz\_4    & idea   & 500m       & 3000        & 500       & mixed\_2 & 1514643752083 & 1514643752609 & FALSE        & 1         & quantitative               & avg       & 0         & 2               & 2            & 0.00            & 0.00              & 0.00          & 0.00             & 0.01        & 0.00          \\
26 & 18          & viz\_5    & idea   & 500m       & 3000        & 500       & mixed\_2 & 1514643752085 & 1514643752616 & FALSE        & 1         & quantitative               & count     & 2         & 9               & 19           & 2.79            & 6.96              & 0.53          & 0.00             & 1.08        & 0.85          \\
27 & 19          & viz\_6    & idea   & 500m       & 3000        & 500       & mixed\_2 & 1514643752637 & 1514643753159 & FALSE        & 1         & quantitative               & count     & 0         & 41              & 60           & 0.23            & 0.74              & 0.32          & 0.00             & 0.37        & 0.52          \\ \hline

\end{tabular} 
}
\end{table*}

Table \ref{tab:detailed_report} shows an example of a detailed report generated by \idebench{}.
Each row describes a query and its evaluation results.
\emph{id} is a query identifier.
\emph{interaction\_id} is a reference to the interaction a query is associated with; it is the index to an interaction in a workflow specification.
\emph{viz\_name} is a reference to the visualization specification in a workflow.
\emph{driver} is the name of the driver used to run the benchmark.
\emph{think\_time}, \emph{time\_req}, \emph{data\_size} refer to the benchmark settings (see Section \ref{sec:settings}).
\emph{workflow} is a the name of the workflow a query is part of.
\emph{start\_time} and \emph{end\_time} are UNIX time stamps of when a query was initiated and when it returns/got cancelled.
\emph{bins\_dims} indicate the number of binning dimensions in the visualization specification.
\emph{binning\_type} indicate whether a nominal and/or quantitative bin range was used in the visualization specification.
\emph{bins\_ofm} is a count of how many bins in the result of a query exceed the margin of error.
\emph{bin\_in\_gt} shows how many bins are in the ground-truth for a query.
\emph{agg\_type} shows which aggregate function was used in the query specification.
The remaining columns are the results for the metrics described in Section \ref{sec:metrics}.
Note that \emph{rel\_error\_avg} and \emph{margin\_avg} are the mean relative error/margin of error across all bins in a query result.
The summary report (see Section \ref{fig:overall_results}) computes the means of all metrics in the detailed report, aggregated on a per workflow-type basis.

\end{appendices}

\begin{scriptsize}
\bibliographystyle{abbrv}
\bibliography{main}
\end{scriptsize}

\end{document}